\documentclass[prd,aps,twocolumn,showpacs,,preprintnumbers,amsmath,amssymb]{revtex4}
\usepackage[dvips]{graphicx}
\usepackage{amsmath}
\usepackage{amssymb}

\usepackage{graphicx}
\usepackage{dcolumn}
\usepackage{bm}
\def\be{\begin{equation}}
\def\ee{\end{equation}}
\def\bea{\begin{eqnarray}}
\def\eea{\end{eqnarray}}

\begin{document}

\preprint{HUTP-06/A0032}

\date{\today}

\title{String Gas Cosmology and Structure Formation}

\author{Robert H. Brandenberger$^1$} \email[email: ]{rhb@hep.physics.mcgill.ca}
\author{Ali Nayeri$^2$} \email[email: ]{nayeri@feynman.harvard.edu}
\author{Subodh P. Patil $^1$} \email[email: ]{patils@hep.physics.mcgill.ca}
\author{Cumrun Vafa $^{2,}$} \email[email: ]{vafa@physics.harvard.edu}

\affiliation{\qquad $^1$~ Department of Physics, McGill University,
Montr\'eal, QC, H3A 2T8, Canada \\
$^2$~Jefferson Physical Laboratory, Harvard University, Cambridge,
MA 02138, U.S.A.}


\begin{abstract}
It has recently been shown that a Hagedorn phase of string gas
cosmology may provide a causal mechanism for generating a nearly
scale-invariant spectrum of scalar metric fluctuations, without the
need for an intervening period of de Sitter expansion. A distinctive
signature of this structure formation scenario would be a slight
blue tilt of the spectrum of gravitational waves. In this paper we
give more details of the computations leading to these results.
\end{abstract}

\maketitle

\newcommand{\eq}[2]{\begin{equation}\label{#1}{#2}\end{equation}}

\section{Introduction}

String gas cosmology (SGC) \cite{BV,TV} (see also \cite{Perlt} for
early work, \cite{rbr1,rbr2} for recent overviews, and \cite{str}
for a critical review) is an approach to string cosmology, based on
adding minimal but crucial inputs from string theory, namely new
degrees of freedom - string oscillatory and winding modes - and new
symmetries - T-duality - to the now standard hypothesis of a hot and
small early universe (see \cite{others} for other work on string gas
cosmology, and in particular \cite{ABE} for a discussion of the role
which D-branes play in this scenario). One key aspect of SGC is that
the temperature cannot exceed a limiting temperature, the Hagedorn
temperature $T_H$ \cite{Hagedorn}. This immediately provides a
qualitative reason which leads us to expect that string theory can
resolve cosmological singularities \cite{BV}. From the equations of
motion of string gas cosmology \cite{TV,Ven} it in fact follows that
if we follow the radiation-dominated Friedmann-Robertson-Walker
(FRW) phase of standard cosmology into the past, a smooth transition
to a quasi-static Hagedorn phase will occur. In this phase, the
string metric is quasi-static while the dilaton is time-dependent.
Reversing the time direction in this argument, we can set up the
following new cosmological scenario \cite{BV}: the universe starts
out in a quasi-static Hagedorn phase during which thermal
equilibrium can be established over a large scale (a scale
sufficiently large for our current universe to grow out of it
following the usual non-inflationary cosmological dynamics). The
quasi-static phase, however, is not a stable fixed point of the
dynamics, and eventually a smooth transition to a
radiation-dominated FRW phase will occur, after which point the
universe evolves as in standard cosmology. This transition is
connected to the decay of string winding modes into string
oscillatory degrees of freedom, as described initially in \cite{BEK}
and studied later in more depth in \cite{Col,Frey}.

Recently \cite{Ali1,Ali3} it was discovered that, under certain
assumptions which will be detailed below, string thermodynamic
fluctuations during the Hagedorn phase lead to scalar metric
fluctuations which are adiabatic and nearly scale-invariant at late
times, thus providing a simple alternative to slow-roll inflation
for establishing such perturbations. It is to be emphasized that
this mechanism for the generation of the primordial perturbations is
intrinsically stringy - particle thermodynamic fluctuations would
lead to a spectrum with a large and phenomenologically unacceptable
blue tilt \cite{Ali3}. As discussed in \cite{Ali2}, the stringy
thermal fluctuations also generate an almost scale-invariant
spectrum of gravitational waves (tensor metric fluctuations).
Whereas the spectrum of scalar modes has a slight red tilt (like the
fluctuations generated in most inflationary models), the
gravitational wave spectrum has a slight blue tilt. This is a
prediction which distinguishes our string gas cosmology scenario
from standard inflationary models. In inflationary models, a slight
red tilt of the spectrum is expected since the amplitude of
gravitational waves is set by $H$, and $H$ is decreasing in time,
leading to lower values of $H$ when scales with larger values of the
comoving wavenumber $k$ exit the Hubble radius.

In this paper, we review the analyses of our previous results
\cite{Ali1,Ali3,Ali2} and present more details. The outline of this
paper is as follows. We begin with a discussion of the background
string gas cosmology, emphasizing the smooth transition between the
initial quasi-static Hagedorn phase and the later
radiation-dominated period. Next, we describe how the scalar and
tensor metric fluctuations are related to correlation functions of
the string gas energy-momentum tensor. In Section 4 we then discuss
the computation of the string gas correlation functions. Sections 5
and 6 describe the evolution of scalar and tensor metric
fluctuations, respectively. In the final section we discuss our
results, point out unresolved issues and future directions for
research.

In the following, we assume that our three spatial dimensions are
already large (sufficiently large such that expansion according to
standard cosmology beginning at a temperature of the string scale
can produce our observed universe today - if the string scale
is of the order $10^{16}$ GeV, then the initial size needs to
be of the order of 1mm) during the Hagedorn phase (for a possible mechanism to
achieve this see \cite{Natalia1}), while the extra spatial
dimensions are confined to the string scale. For a mechanism to
achieve the separation into three large and six small compact dimensions
in the context of SGC see \cite{BV} (see however
\cite{Col,Frey} for caveats), and for a natural dynamical mechanism
arising from SGC to stabilize all of the moduli associated with the
extra spatial dimensions see
\cite{Watson,Patil1,Patil2,Patil3,Edna,shiggs} (see also
\cite{other2}).

To be specific, we
take our three dimensions to be toroidal. The existence of one
cycles results in the stability of string winding modes - and this
is a key ingredient in our calculations.

\section{The Background}

String gas cosmology \cite{BV,TV} is based on coupling an ideal gas of
strings to a dilaton-gravity background, and on focusing on
the consequences which follow from the existence of new
stringy degrees of freedom and new stringy symmetries, namely
degrees of freedom and symmetries which are not present in
point particle theories.

In the case of a homogeneous
and isotropic background given by
\be
ds^2 \, = \, dt^2 - a(t)^2 d{\bf x}^2 \, ,
\ee
the three resulting equations of motion of dilaton-gravity (the
generalization of the two Friedmann equations plus the equation
for the dilaton) in the string frame are
\cite{TV} (see also \cite{Ven})
\bea
-(d) {\dot \lambda}^2 + {\dot \varphi}^2 \, &=& \, e^{\varphi}
E
\label{E1} \\
{\ddot \lambda} - {\dot \varphi} {\dot \lambda} \, &=& \,
{1 \over 2} e^{\varphi} P \label{E2} \\
{\ddot \varphi} - (d) {\dot \lambda}^2 \, &=& \, {1 \over 2}
e^{\varphi} E \, , \label{E3} \eea
where $E$ and $P$ denote the total energy and pressure,
respectively, $d$ is the number of spatial dimensions,
and we have introduced the logarithm of the scale factor
\be \lambda(t) \, = \, \ln{ [a(t)]} \,, \ee
and the rescaled dilaton
\be
\varphi \, = \, 2 \phi - (d) \lambda \, .
\ee

If we take the spatial topology to be that of a torus with radius $R$,
the new stringy degrees of freedom are string winding modes whose
energies are quantized in units of $R$, and oscillatory modes, whose
energies are independent of $R$. The number of oscillatory modes
of a string increases exponentially with energy, thus leading
to a maximal temperature (the Hagedorn temperature $T_H$ \cite{Hagedorn})
for a gas of strings.

Since the energies of the
momentum modes of a string are quantized in units of $l_s^2/R$,
where $l_s$ is the string length, the spectrum of mass states of the
free string is symmetric under the interchange $R \rightarrow l_s^2 / R$,
the T-duality transformation. String vertex operators are also invariant
under this transformation, and assuming that the symmetry extends to
non-perturbative string theory leads to the existence of D-branes as
new fundamental objects in string theory \cite{Pol}.

In thermal equilibrium at the string scale ($R \simeq l_s$), the
self-dual radius, the number of winding and momentum modes will be
equal. Since winding and momentum modes give an opposite
contribution to the pressure, the pressure of the string gas in
thermal equilibrium at the self-dual radius will vanish. From the
above equations of motion it then follows that a static phase
$\lambda = 0$ will be a fixed point of the dynamical system. This
phase is called the ``Hagedorn phase''.

On the other hand, for large values of $R$ in thermal equilibrium
the energy will be exclusively in momentum modes. These act as usual
radiation. Inserting the radiative equation of state into the above
equations (\ref{E1} - \ref{E3}) it follows that the source in
the dilaton equation of motion
\be
{\ddot \phi} + (d) {\dot \lambda} {\dot \phi} - 2 {\dot \phi}^2
\, = \, - {1 \over 2} e^{\varphi} E [1 - (d) w] \, ,
\ee
where $w = P/E$ is the equation of state parameter,
vanishes and the dilaton approaches
a constant as a consequence of Hubble damping. Consequently, the
scale factor expands as in the usual radiation-dominated universe.

The transition between the Hagedorn phase and the radiation-dominated
phase with fixed dilaton is achieved via the annihilation of winding
modes, as studied in detail in \cite{BEK}.
The main point is that,
starting in a Hagedorn phase, there will be a smooth transition to
the radiation-dominated phase of standard cosmology with fixed dilaton.
It is in this cosmological background that we will study the
generation of fluctuations.

It is instructive to compare the background evolution of string gas
cosmology with the background of inflationary cosmology
\cite{Guth,Sato}, the current
paradigm of early Universe evolution. Figure 1 is a sketch of
the space-time evolution in string gas cosmology. For times $t < t_R$,
we are in the quasi-static Hagedorn phase, for $t > t_R$ we have
the radiation-dominated period of standard cosmology. To understand
why string gas cosmology can lead to a causal mechanism of structure
formation, we must compare the evolution of the physical
wavelength corresponding to a fixed comoving scale (fluctuations in
early universe cosmology correspond to waves with a fixed wavelength
in comoving coordinates) with that of the Hubble radius $H^{-1}(t)$,
where $H(t)$ is the expansion rate. The Hubble radius separates
scales on which fluctuations oscillate (wavelengths smaller than
the Hubble radius) from wavelengths where the fluctuations are frozen
in and cannot be effected by microphysics. Causal microphysical
processes can generate fluctuations only on sub-Hubble scales
(see e.g. \cite{RHBrev} for
a concise overview of the theory of cosmological perturbations and
\cite{MFB} for a comprehensive review).

\begin{figure}
\includegraphics[height=10cm]{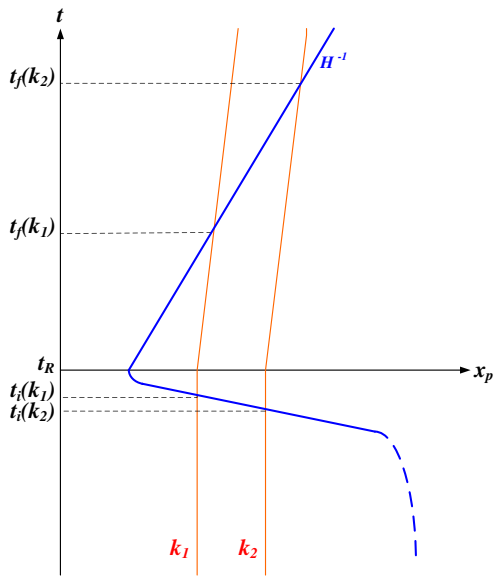}
\caption{Space-time diagram (sketch) showing the evolution of fixed
comoving scales in string gas cosmology. The vertical axis is time,
the horizontal axis is physical distance. The Hagedorn phase ends at
the time $t_R$ and is followed by the radiation-dominated phase of
standard cosmology. The solid curve represents the Hubble radius
$H^{-1}$ which is cosmological during the quasi-static Hagedorn
phase, shrinks abruptly to a microphysical scale at $t_R$ and then
increases linearly in time for $t > t_R$. Fixed comoving scales (the
dotted lines labeled by $k_1$ and $k_2$) which are currently probed
in cosmological observations have wavelengths which are smaller than
the Hubble radius during the Hagedorn phase. They exit the Hubble
radius at times $t_i(k)$ just prior to $t_R$, and propagate with a
wavelength larger than the Hubble radius until they reenter the
Hubble radius at times $t_f(k)$.} \label{fig:1}
\end{figure}

In string gas cosmology, the Hubble radius is infinite deep in the
Hagedorn phase. As the universe starts expanding near $t_R$, the
Hubble radius rapidly decreases to a microscopic value (set by the
temperature at $t = t_R$ which will be close to the Hagedorn
temperature), before turning around and increasing linearly in the
post-Hagedorn phase. The physical wavelength corresponding to a
fixed comoving scale, on the other hand, is constant during the
Hagedorn era. Thus, all scales on which current experiments measure
fluctuations are sub-Hubble deep in the Hagedorn phase. In the
radiation period, the physical wavelength of a perturbation mode
grows only as $\sqrt{t}$. Thus, at a late time $t_{f}(k)$ the
fluctuation mode will re-enter the Hubble radius, leading to the
perturbations which are observed today.

In contrast, in inflationary cosmology (Figure 2) the Hubble radius
is constant during inflation ($t < t_R$, where here $t_R$ is the
time of inflationary reheating), whereas the physical wavelength
corresponding to a fixed comoving scale expands exponentially. Thus,
as long as the period of inflation is sufficiently long, all scales
of interest for current cosmological observations are sub-Hubble at
the beginning of inflation.

\begin{figure}
\includegraphics[height=10cm]{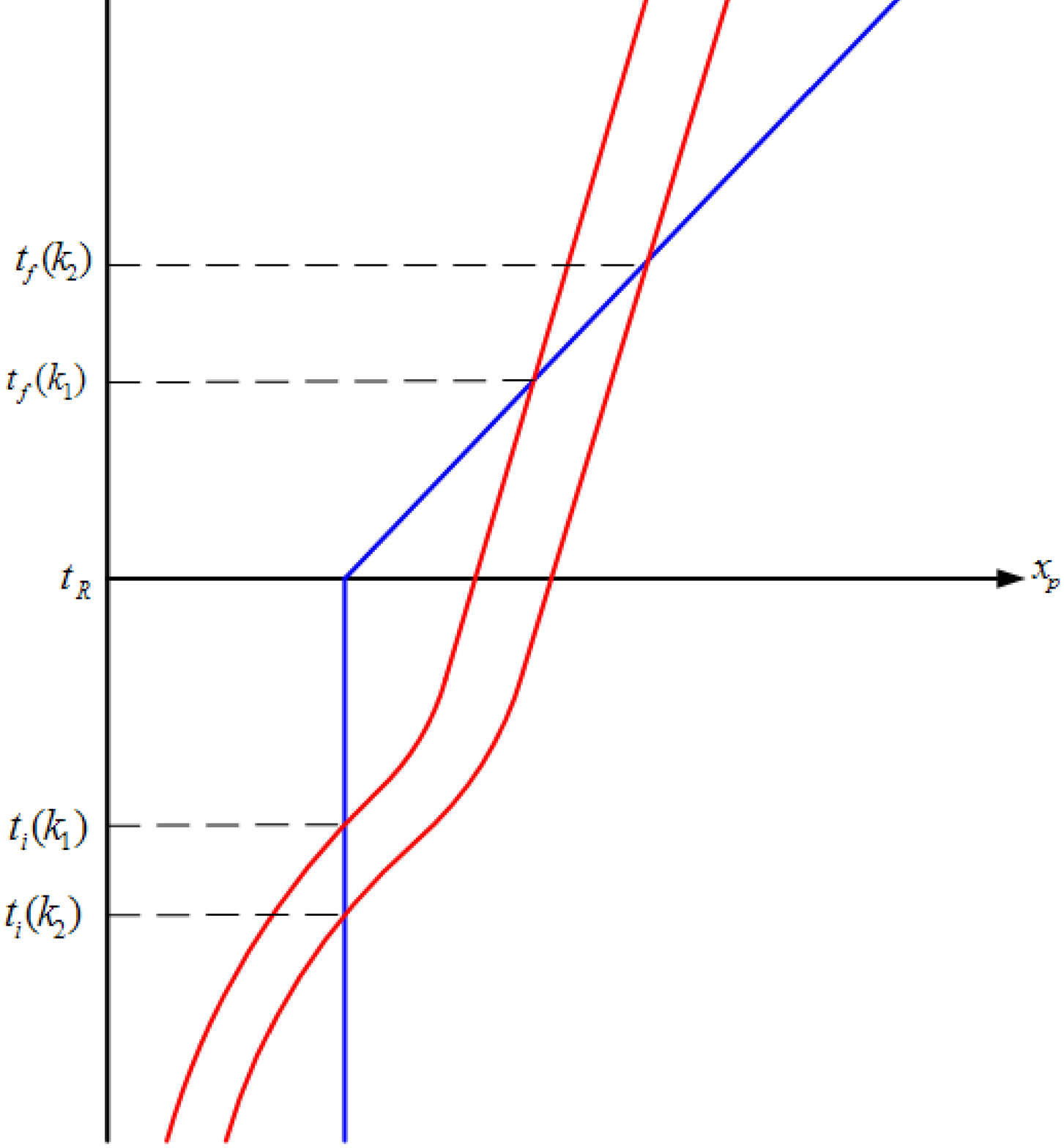}
\begin{caption}
{\small Corresponding space-time diagram (sketch) for inflationary
cosmology. The period of exponential expansion is for $t < t_R$. At
the time of reheating $t = t_R$ the universe makes a transition to a
radiation-dominated phase. As in Figure 1, the vertical axis is time
and the horizontal axis is physical distance. The solid curve
represents the Hubble radius. Fixed comoving scales (the dotted
lines labeled by $k_1$ and $k_2$) which are currently probed in
cosmological observations have wavelengths which start out smaller
than the Hubble radius at the beginning of the inflationary phase.
They exit the Hubble radius at times $t_i(k)$ and propagate with a
wavelength larger than the Hubble radius until they reenter the
Hubble radius at times $t_f(k)$.}
\end{caption}
\label{fig:2}
\end{figure}

In spite of the fact that both in inflationary cosmology and in string
gas cosmology, scales are sub-Hubble during the early stages, the
actual generation mechanism for fluctuations is completely different.
In inflationary cosmology, any thermal fluctuations present before
the onset of inflation are red-shifted away, leaving us with a quantum
vacuum state, whereas in the quasi-static Hagedorn phase of string gas
cosmology matter is in a thermal state. Hence, whereas in inflationary
cosmology the fluctuations originate as quantum vacuum perturbations
\cite{Mukh,Press},
in string gas cosmology the inhomogeneities are created by the thermal
fluctuations of the string gas.

As we have shown in \cite{Ali1,Ali3,Ali2}, string thermodynamics in
the Hagedorn phase of string gas cosmology yields an almost
scale-invariant spectrum of both scalar and tensor modes. This is an
intrinsically stringy result: thermal fluctuations of a gas of
particles would lead to a very different (namely very ``blue''
spectrum). Since the primordial perturbations in our scenario are of
thermal origin (and there are no non-vanishing chemical potentials),
they will be adiabatic.

In the following, we will discuss the derivation of these results in
more details than discussed in our initial papers \cite{Ali1,Ali2}.

\section{Scalar and Tensor Metric Fluctuations}

In this section, we show how the scalar and tensor metric fluctuations
can be extracted from knowledge of the energy-momentum tensor of
the string gas.

Working in conformal time $\eta$ (defined via
$dt = a(t)d\eta$), the metric of a homogeneous and isotropic background
space-time perturbed by linear scalar metric fluctuations and
gravitational waves can be written in the form
\be
d s^2 = a^2(\eta) \left\{(1 + \Phi)d\eta^2 - [(1 -
\Psi)\delta_{ij} + h_{ij}]d x^i d x^j\right\} \,.
\ee
Here, $\Phi$ and $\Psi$ describe the scalar metric fluctuations.
Both are functions of space and time. The tensor $h_{ij}$ is
transverse and traceless and contains the two polarization states of
the gravitational waves. In the above, we have fixed the coordinate
freedom by working in the so-called ``longitudinal" gauge in which
the scalar metric fluctuation is diagonal. Note that to linear order
in the amplitude of the fluctuations, scalar and tensor modes
decouple, and the tensor modes are gauge-invariant. Note also that
in the absence of non-adiabatic stress, the two degrees of freedom
$\Phi$ and $\Psi$ for scalar metric fluctuations coincide (see
\cite{MFB} for a comprehensive review of the theory of cosmological
perturbations and \cite{RHBrev} for a pedagogical introduction).

Inserting the above metric into the Einstein equations, subtracting
the background terms and truncating the perturbative expansion at linear
order leads to the following system of equations
\begin{eqnarray} \label{perteom1}
- 3 {\cal H} \left( {\cal H} \Phi + \Psi^{'} \right) + \nabla^2 \Psi
\,
&=& \, 4 \pi G a^2 \delta T^0{}_0 \nonumber \\
\left( {\cal H} \Phi + \Psi^{'} \right)_{, i} \,
&=& 4 \pi G a^2 \delta T^0{}_i  \nonumber \\
\left[ \left( 2 {\cal H}^{'} + {\cal H}^2 \right) \Phi + {\cal H}
\Phi^{'}
+ \Psi^{''} + 2 {\cal H} \Psi^{'} \right] && \nonumber \\
+ {1 \over 2} \nabla^2 D  - {1 \over 2} D_{,i}^i \,
&=& - 4 \pi G a^2 \delta T^i{}_i \, , \nonumber \\
-{1 \over 2} \left[ {{a^{\prime \prime}} \over a} - {1 \over
2}\left({{a^{\prime}} \over a}\right)^2 \right] h_{ij}
+ {1 \over 4} {{a^{\prime}} \over a} h_{ij}^{\prime} && \nonumber \\
+ \left[{{\partial^2} \over {\partial \eta^2}} - \nabla^2\right]
h_{ij} + {1 \over 2} D_{,ij} \,
&=& - 4 \pi G a^2 \delta T^i{}_j \,,\nonumber \\
&& \mbox{for $i \neq j$}\,,
\end{eqnarray}
where $D = \Phi - \Psi$, ${\cal H} = a^{\prime} / a$, a prime
denotes the derivative with respect to conformal time $\eta$,
and $G$ is Newton's gravitational constant.

In the Hagedorn phase, these equations simplify substantially and
allow us to extract the scalar and tensor metric fluctuations
individually. First of all, since there is no anisotropic stress at
late times, $D = 0$ and hence $\Phi = \Psi$. Replacing comoving by
physical coordinates, we obtain from the $00$ equation
\be
\label{scalar} \nabla^2 \Phi \, = \, 4 \pi G \delta T^0{}_0
\,
\ee
and from the $i \neq j$ equation
\be
\label{tensor} \nabla^2 h_{ij} \, = \, - 4 \pi G \delta T^i{}_j
\, .
\ee

The above equations (\ref{scalar}) and (\ref{tensor}) allow
us to compute the power spectrum of scalar and tensor metric
fluctuations in terms of correlation functions of the
energy-momentum tensor. Since the metric perturbations are
small in amplitude we can consistently work in Fourier space.

Our approximation scheme for computing the cosmological
perturbations and gravitational wave spectra from string gas
cosmology is as follows (the analysis is similar to how
the calculations were performed in \cite{BST,BK} in the
case of inflationary cosmology). For a fixed comoving scale $k$ we follow
the matter fluctuations until the time $t_i(k)$
shortly before the end of the Hagedorn phase when the scale exits
the Hubble radius. At that time, we use (\ref{scalar}) and
(\ref{tensor}) to compute the values of $\Phi(k)$ and $h(k)$ ($h$ is
the amplitude of the gravitational wave tensor), and, since we
are working in Fourier space, we replace the $\nabla^2$ operator by
$k^2$.

The expectation value of the metric fluctuations is thus given in
terms of correlation functions of the string energy-momentum tensor.
Specifically,
\be \label{scalarexp}
\langle|\Phi(k)|^2\rangle \, = \, 16 \pi^2 G^2
k^{-4} \langle\delta T^0{}_0(k) \delta T^0{}_0(k)\rangle \, ,
\ee
where the pointed brackets indicate expectation values, and
\be
\label{tensorexp} \langle|h(k)|^2\rangle \, = \, 16 \pi^2 G^2
k^{-4} \langle\delta T^i{}_j(k) \delta T^i{}_j(k)\rangle \,,
\ee
where on the right hand side of (\ref{tensorexp}) we mean the
average over the correlation functions with $i \neq j$.

After the modes exit the Hubble radius, we follow the evolution of
both scalar and tensor fluctuations by means of the usual equations
of cosmological perturbation theory. Apart from small corrections which
arise in the time interval $t \in [t_i(k), t_R]$, this technique is
justified since the dilaton is fixed after the end of the Hagedorn phase.

In the following section, we describe in a bit more detail than in
our previous letters how to compute the string matter correlation
functions from string thermodynamics. After that, we return to the
computation of the spectrum of scalar and tensor modes.

\section{Energy-Momentum Tensor Correlation Functions for Closed Strings}

In the following we will denote the space-time metric by $g_{\mu \nu}$.
Unlike what was done in \cite{Ali2}, we will in this paper be working with
a Minkowski signature metric.
The starting point for our considerations is the free energy $F$
of a string gas in thermal equilibrium
\be \label{free}
F \, = \,- \frac{1}{\beta} ln Z \, ,
\ee
where $\beta$ is the inverse temperature and the canonical partition
function $Z$ is given by
\be \label{part}
Z \, = \, \sum_s e^{-\beta\sqrt{-g_{00}}H(s)} \, ,
\ee
where the sum runs over the states $s$ of the string gas, and $H(s)$ is
the energy of the state. The action $S$ of the string gas is given
in terms of the free energy $F$ via
\be \label{action}
S \, = \, \int dt \sqrt{-g_{00}} F[g_{ij},\beta] \, .
\ee
Note that the free energy depends on the spatial components of the
metric because the energy of the individual string states does.
The energy-momentum tensor $T^{\mu \nu}$ of the string gas is
determined by varying the action with respect to the metric:
\be \label{emtensor}
T^{\mu\nu} \, = \, \frac{2}{\sqrt{-g}}\frac{\delta S}{\delta g_{\mu\nu}} \, .
\ee

Consider now the thermal expectation value
\be \label{texp1}
\langle T^\mu{}_\nu \rangle \, = \,
\frac{1}{Z}\sum_s T^\mu{}_\nu(s)e^{-\beta \sqrt{-g_{00}}H(s)} \, ,
\ee
where $T^\mu{}_\nu (s)$ and $H(s)$ are the energy momentum tensor
and the energy of the state labeled by $s$, respectively. Making use
of (\ref{emtensor}) and (\ref{action}) we immediately find that
\be
T^\mu{}_\nu (s) \, = \, 2 \frac{g^{\mu \lambda}}{\sqrt{-g}}
\frac{\delta }{\delta g^{\lambda \mu}} [-\sqrt{-g_{00}}H(s)] \,,
\ee
and hence
\be \label{texp2}
\langle T^\mu{}_\nu \rangle \, = \, 2 \frac{g^{\mu
\lambda}}{\sqrt{-g}} \frac{\delta \ln{Z}}{\delta g^{\nu\lambda}} \,
.
\ee

To extract the fluctuation tensor of $T_{\mu \nu}$ for long wavelength modes, we take one
additional variational derivative of (\ref{texp2}), using (\ref{texp1}) to obtain
\bea \label{fluct} \langle T^\mu{}_\nu T^\sigma{}_\lambda \rangle &
- &\langle T^\mu{}_\nu \rangle \langle T^\sigma{}_\lambda \rangle =
2 \frac{g^{\mu \alpha}}{\sqrt{ - g}}\frac{\partial}{\partial
g^{\alpha\nu}}\left(\frac{g^{\sigma \delta}}{\sqrt{ -
g}}\frac{\partial \ln{Z}}{\partial g^{\delta\lambda}}\right)\nonumber \\
&& + 2 \frac{g^{\sigma \alpha}}{\sqrt{ - g}}\frac{\partial}{\partial
g^{\alpha\lambda}}\left(\frac{G^{\mu \delta}}{\sqrt{ -
g}}\frac{\partial \ln{Z}}{\partial g^{\delta\nu}}\right)\,. \eea

As we saw in the previous section, the scalar metric fluctuations are
determined by the energy density correlation function
\be
C^0{}_0{}^0{}_0 \, = \, \langle \delta\rho^2 \rangle \, = \,
\langle \rho^2 \rangle - \langle \rho \rangle ^2 \, .
\ee
We will read off the result from the expression (\ref{fluct})
evaluated for $\mu = \nu = \sigma = \lambda = 0$.
The derivative with respect to $g_{00}$ can be expressed in terms of
the derivative with respect to $\beta$. After a couple of steps of
algebra we obtain 
\be \label{cor1}
C^0{}_0{}^0{}_0 \,
= \, - \frac{1}{R^{6}} \frac{\partial}{\partial \beta}
\left(F + \beta \frac{\partial F}{\partial \beta}\right) \, = \,
\frac{T^2}{R^6} C_V \, .
\ee
where
\be \label{specheat} C_V \, = \, (\partial  E  /
\partial T)|_{V} \,,
\ee
is the specific heat, and
\be
E \, \equiv \, F + \beta \left(\frac{\partial F}{\partial
\beta}\right) \,,
\ee
is the internal energy. Also, $V = R^3$ is the
volume of the three compact but large spatial dimensions.

The gravitational waves are determined by the off-diagonal
spatial components of the correlation function tensor, i.e.
\be \label{cor2a} C^i{}_j{}^i{}_j \, = \, \langle \delta {T^i{}_j}^2
\rangle \, = \, \langle {T^i{}_j}^2 \rangle - \langle T^i{}_j
\rangle^2\,,
\ee
with $i \neq j$.

Our aim is to calculate the fluctuations of the energy-momentum
tensor on various length scales $R$. For each value of $R$, we will
consider string thermodynamics in a box in which all edge lengths
are $R$. From (\ref{fluct}) it is obvious that in order to have
non-vanishing off-diagonal spatial correlation functions, we must
consisted a twisted torus. Let us focus on the $x-y$ component of
the correlation function. We will consider the spatial part of the
metric to be 
\be \label{metric}
ds^2 \, = \, R^2 d\theta_x^2 + 2 \epsilon R^2 d\theta_x d\theta_y
+ R^2 d\theta_y^2
\ee
with $\epsilon \ll 1$. The spatial coordinates $\theta_i$ run over a
fixed interval, e.g. $[0, 2\pi]$), The generalization of the spatial
part of the metric to three dimensions is obvious. At the end of the
computations, we will set $\epsilon = 0$.

{F}rom the form of (\ref{fluct}), it follows that all
space-space correlation function tensor elements are of the same
order of the magnitude, namely 
\be \label{cor2}
C^i{}_j{}^i{}_j
\, = \, \frac{1}{\beta R^3}\frac{\partial}{\partial \ln{R}}\left(-
\frac{1}{R^3} \frac{\partial F}{\partial \ln{R}}\right)  =
\frac{1}{\beta R^2}\frac{\partial p}{\partial R} \, ,
\ee
where the string pressure is given by
\be
p \,  \equiv   \, -\frac{1}{V}\left(\frac{\partial F}{\partial
\ln{R}}\right) \, = \, T \left(\frac{\partial S}{\partial
V}\right)_E \,. \label{stringypressure}
\ee

In the following, we will compute the two correlation functions
(\ref{cor1}) and (\ref{cor2}) using tools from string statistical
mechanics. Specifically, we will be following the discussion in
\cite{Deo} (see also \cite{Deo1,Deo2,mb,mt}). The starting point is
the formula
\be
S(E , R) \, = \, \ln{\Omega(E ,R)}
\ee
for the entropy in terms of $\Omega(E ,R)$, the density of states.
The density of states of a gas of closed strings on a large
three-dimensional torus (with the radii of all internal dimensions
at the string scale) was calculated in \cite{Deo} (see also
\cite{Ali3}) and is given by
\be
\Omega(E , R) \, \simeq \, \beta_H e^{\beta_H E + n_H V}[1 +
\delta \Omega_{(1)}(E , R)] \label{density_states}\,,
\ee
where $\delta \Omega_{(1)}$ comes from the contribution to the
density of states (when writing the density of states as a Laplace
transform of $Z(\beta)$, which involves integration over $\beta$)
from the closest singularity point $\beta_1$ to $\beta_H = (1/T_H)$
in the complex $\beta$ plane. Note that $\beta_1 < \beta_H$, and
$\beta_1$ is real. From \cite{Deo,Ali3} we have
\be \label{deltaomega}
\delta \Omega_{(1)}(E , R) \, = \, - \frac{(\beta_H E)^{5}}{5!}
e^{-(\beta_H - \beta_1)(E  - \rho_H V)} \,.
\ee
In the above, $n_H$ is a (constant) number density of order
$l_s^{-3}$  and $\rho_H$ is the `Hagedorn Energy density' of the
order $l_s^{-4}$, and
\be
\beta_H - \beta_1 \sim \left\{ \begin{array}{ll}
(l_s^3/R^2) \,, & \mbox{for $R \gg l_s$}\,, \\
(R^2/l_s)\,, & \mbox{for $R \ll l_s$}\,.
\end{array}
\right.
\ee
To ensure the validity of Eq. (\ref{density_states}) we demand that
\be \label{cond2}
- \delta \Omega_{(1)} \,\ll \, 1
\ee
by assuming $\rho \equiv (E / V) \gg \rho_H$.

Combining the above results, we find that the entropy of the string gas
in the Hagedorn phase is given by
\be
\label{entropy} S(E , R) \simeq \beta_H E + n_H V + \ln{\left[1
+ \delta \Omega_{(1)}\right]} \,,
\ee
and therefore the temperature $T (E, R) \equiv [(\partial S/\partial
E)_V]^{-1}$ will be
\bea
T  &\simeq&  \left(\beta_H + \frac{\partial \delta
\Omega_{(1)}/\partial E}{1 + \delta \Omega_{(1)}}\right)^{-1} \nonumber \\
&\simeq&
T_H \left(1 + \frac{\beta_H - \beta_1}{\beta_H} \delta
\Omega_{(1)}\right)\label{temp}\,.
\eea
In the above, we have dropped a term which is negligible since
$E (\beta_H - \beta_1) \gg 1$ (see (\ref{cond2})).
Using this relation, we can express $\delta \Omega_{(1)}$ in terms of
$T$ and $R$ via
\be \label{deltaomega}
l_s^3\delta \Omega_{(1)} \, \simeq \, -
\frac{R^2}{T_H}\left(1 - \frac{T}{T_H}\right) \, .
\ee
In addition, we find
\be \label{meanen}
E  \, \simeq \,  l_s^{-3} R^2
\ln{\left[\frac{\ell_s^3 T}{R^2 (1- T/T_H)}\right]}\,.
\ee
Note that to ensure that $|\delta \Omega_{(1)}| \ll 1$ and $E \gg
\rho_H R^3$, one should demand 
\be \label{cond}
(1 - T/T_H) R^2 l_s^{-2} \ll 1 \, .
\ee

The results (\ref{entropy}) and (\ref{deltaomega}) now allow us
to compute the correlation functions (\ref{cor1}) and (\ref{cor2}).
We first compute the energy correlation function (\ref{cor1}).
Making use of (\ref{meanen}), it follows from
(\ref{specheat}) that
\be \label{specheat2}
C_V  \, \approx \,  \frac{R^2/l_s^3}{T \left(1
- T/T_H\right)} \,,
\ee
from which we get
\be C^0{}_0{}^0{}_0 = \langle\delta \rho^2\rangle\, \simeq \,
\frac{T}{l_s^3(1 - T/T_H)} \frac{1}{R^4}\, . \ee
Note that the factor $(1 - T/T_H)$ in the denominator is responsible
for giving the spectrum a slight red tilt. It comes from the differentiation
with respect to $T$.

Next we evaluate (\ref{stringypressure}). From the definition of the
pressure it follows that (to linear order in $\delta \Omega_{(1)}$)
\be
p \, = \, n_H T + T {{\partial} \over {\partial V}} \delta \Omega_{(1)}
\, ,
\ee
where the final partial derivative is to be taken at constant
energy. In taking this partial derivative, we insert the
expression (\ref{deltaomega}) for $\delta \Omega_{(1)}$ and must keep
careful account of the fact that $\beta_H - \beta_1$ depends
on the radius $R$. In evaluating the resulting terms, we
keep only the one which dominates at high energy density. It
is the term which comes from differentiating the factor
$\beta_H - \beta_1$. This differentiation brings down a
factor of $E$, which is then substituted by means
of (\ref{meanen}), thus introducing a logarithmic factor
in the final result for the pressure. We obtain
\be
p(E, R) \approx n_H T_H - \frac{2}{3}\frac{(1 - T/T_H)}{l_s^3
R}\ln{\left[\frac{l_s^3 T}{R^2 (1- T/T_H)}\right]} \,,
\ee
which immediately yields
\be
\label{tensorresult} C^i{}_j{}^i{}_j \, \simeq \, \frac{T (1 -
T/T_H)}{l_s^3 R^4} \ln^2{\left[\frac{R^2}{l_s^2}(1 -
T/T_H)\right]}\, .
\ee
Note that since no temperature derivative is taken, the factor $(1 - T/T_H)$
remains in the numerator. This is one of the two facts which leads to the
slight blue tilt of the spectrum of gravitational waves. The second
factor contributing to the slight blue tilt is the explicit factor
of $R^2$ in the logarithm. Because of (\ref{cond}), we are on the
large $k$ side of the zero of the logarithm. Hence, the greater the
value of $k$, the larger the absolute value of the logarithmic factor.

\section{Power Spectrum of Scalar Metric Fluctuations}

Let us recall the philosophy behind our computation of the spectrum
of scalar metric fluctuations. For a fixed comoving scale labeled by
its wavenumber $k$, we follow the energy-momentum tensor of the
string matter fluctuations until the time $t_i(k)$ when the scale
exits the Hubble radius. At that point, we use the Einstein
constraint equation (\ref{scalarexp}) to determine the corresponding
metric fluctuations. On super-Hubble scales for $t > t_i(k)$ we use
the usual equations for cosmological perturbations to evolve the
fluctuations.

The dimensionless power spectrum of the cosmological perturbations
(scalar metric fluctuations) is defined by (see e.g. \cite{Peebles})
\be
P_{\Phi}(k) \, \equiv \, k^3 |\Phi(k)|^2 \,
\ee
where we are using the convention of defining the Fourier
transform of a function $f(x)$ including a factor of the
square root of the volume of space. Thus, the dimension of
$\Phi(k)$ is $k^{-3/2}$.

Making use of (\ref{scalarexp}) to replace the expectation
value of $|\Phi(k)|^2$ in terms of the correlation function
of the energy density, we obtain
\be \label{power1}
P_{\Phi}(k) \, = \, 16 \pi^2 G^2 k^{-1}
\langle|\delta \rho(k)|^2\rangle \, .
\ee
The Fourier space expectation value of $\delta \rho$ is related to
the position space root mean square density contrast $\delta M$ in a
region of radius $R = k^{-1}$ via
\be
\langle(\delta M)^2\rangle_R \, = \, k^{-3} \langle|\delta
\rho(k)|^2\rangle \, ,
\ee
and hence (\ref{power1}) becomes
\bea \label{power2}
P_{\Phi}(k) \, &=& \, 16 \pi^2 G^2 k^2 \langle(\delta M)^2\rangle_R \\
               &=& \, 16 \pi^2 G^2 k^{-4} \langle(\delta \rho)^2\rangle_R
\, , \nonumber
\eea
where in the second step we have replaced the mass
fluctuation by the position-space density fluctuation.
According to (\ref{cor1}), the density correlation function
is given by the specific heat via $T^2 R^{-6} C_V$, and hence
(\ref{power2}) becomes
\be \label{power3}
P_{\Phi}(k) \, = \, 16 \pi^2 G^2 k^2 T^2 C_V \, .
\ee
Inserting the expression from (\ref{specheat2}) for the specific
heat of a string gas on a scale $R$ yields our final
result
\be
\label{power4} P_{\Phi}(k) \, = \, 16 \pi^2 G^2 {T \over
{l_s^3}} {1 \over {1 - T/T_H}}
\ee
for the power spectrum of cosmological fluctuations.

During
the Hagedorn phase, the temperature $T$ is close to the Hagedorn
temperature $T_H$, which in turn is given by the string scale.
Thus, to a first approximation, when evaluating the spectrum
(\ref{power4}) at the time $t_i(k)$, the temperature $T(t_i(k))$
is independent of $k$. Thus, the spectrum of scalar metric
fluctuations is scale-invariant, with an amplitude given
by
\be \label{scalarspectrum} P_{\Phi}(k) \, \sim \, \left( {{l_{pl}}
\over {l_s}} \right)^4 {1 \over {1 - T(t_i(k))/T_H}} \, . \ee 
The last factor yields a small red tilt of the spectrum since
$T(t_i(k))$ is a slowly decreasing function of $k$ (modes with
larger values of $k$ are exiting the Hubble radius later, i.e.
at a slightly lower value of $T(t_i(k))$).

As long as the time-dependence of the dilaton is negligible, then
on super-Hubble scales the quantity $\zeta$ which corresponds
to the curvature perturbation in comoving gauge is constant (modulo
effects which come from the decaying mode of $\Phi$ at $t_i(k)$
\cite{Liddle}). The quantity $\zeta$ is defined by \cite{BST,BK,Lyth}
\be
\zeta \, \equiv \, \Phi + {2 \over 3}
{{\bigl(H^{-1} {\dot \Phi} + \Phi \bigr)} \over { 1 + w}} \, ,
\ee
where $w = p / \rho$ is the equation of state parameter.

Unlike in inflationary cosmology, where the factor $1 + w$ changes
by many orders of magnitude during reheating, in string gas
cosmology $1 + w$ makes a very mild transition between an initial
value of $(1 + w)|_i = 1$ and a final value of $(1 + w)|_f = 4/3$
between the Hagedorn phase and the radiation-dominated phase. Thus,
up to a factor of order unity, the conservation of $\zeta$ on
super-Hubble scales in our case leads to the conservation of $\Phi$.

It is important to realize, however, that in spite of the fact that
the amplitude of $\Phi$ is constant
on super-Hubble scales, the fluctuation mode is evolving non-trivially -
it is being highly squeezed by the expansion of the background. One
way to see this is to realize that when the fluctuations cross the
Hubble radius, ${\dot \Phi}$ is comparable to $k \Phi$ (since the
fluctuations are thermal in origin). However, for $t > t_i(k)$
the amplitude of ${\dot \Phi}$ decreases proportionally to $a^{-1}(t)$.
The fluctuation mode approaches the form of a standing mode at rest.
This squeezing is analogous to what happens in inflationary cosmology.
The only difference is that in inflationary cosmology, the squeezing
takes place both during and after inflation, whereas in string gas
cosmology the squeezing happens only during the period of
radiation-domination.

Another way to understand the squeezing of cosmological fluctuations
is in terms of canonical variables (variables characterizing the
cosmological fluctuations in terms of which the action for the
perturbations has canonical kinetic term). One of the key
results of the quantum theory of cosmological fluctuations
\cite{MFB,RHBrev} is
that the variable $v$ in terms of which the action for fluctuations
(for ideal gas matter) has the canonical form
\be \label{pertaction}
S \, = \, {1 \over 2} \int d^4x \left(
{v^{\prime}}^2 - c_s^2 v_{,i}v_{,i} + {{z^{\prime \prime}} \over z}
v^2 \right)
\ee
is the so-called Sasaki-Mukhanov variable \cite{Sasaki, Mukh2}
\be \label{Mukhvar}
v \, \equiv \, {1 \over {\sqrt{6}l_s}} \left(
\varphi_v - 2 z \Phi \right) \, ,
\ee
where $\phi_v$ is the velocity potential of the ideal fluid, $c_s$ is
the speed of sound, and $z$ is a function of the background given by
\bea
z(\eta) \, &=& \, a(\eta)
{{\sqrt{{\cal{H}}^2 - {\cal H}^{\prime}}} \over {c_s {\cal H}}} \nonumber \\
& \propto & \, a(\eta) \, .
\eea
The action (\ref{pertaction}) is that of a free scalar field with a
time-dependent negative square mass. The mass term dominates on
super-Hubble scales and leads to the squeezing of the wave function
for $v$, squeezing which is proportional to $a$.

The squeezing of fluctuations is crucial in order to obtain the
``acoustic'' oscillations in the angular power spectrum of the
cosmic microwave background (CMB) anisotropies \cite{SZ}. If all
modes enter the Hubble radius at late times as standing waves and
then begin to oscillate, then at the time of last scattering
$t_{LS}$ the wave which is entering the Hubble radius at that time
will lead to maximal redshift fluctuations, whereas a mode which has
entered the Hubble radius earlier and has had time to perform $1/4$
oscillation will have zero amplitude and maximal velocity at
$t_{LS}$, and on the corresponding angular scale the redshift
contribution to the CMB anisotropies will vanish. Models without
squeezing of fluctuations on super-Hubble scales do not yield this
pattern of acoustic oscillations, even if they, such as in the case
of topological defect models of structure formation \cite{CS},
generate a scale-invariant spectrum of fluctuations.

\section{Power Spectrum of Tensor Metric Fluctuations}

In this section we
estimate the dimensionless power spectrum of gravitational waves.
First, we make some general comments. In slow-roll inflation, to
leading order in perturbation theory matter fluctuations do not
couple to tensor modes. This is due to the fact that the matter
background field is slowly evolving in time and the leading order
gravitational fluctuations are linear in the matter fluctuations. In
our case, the background is not evolving (at least at the level of
our computations), and hence the dominant metric fluctuations are
quadratic in the matter field fluctuations. At this level, matter
fluctuations induce both scalar and tensor metric fluctuations.
Based on this consideration we expect that in our string gas
cosmology scenario, the ratio of tensor to scalar metric
fluctuations will be larger than in simple slow-roll inflationary
models.

The method for calculating the spectrum of gravitational waves
is similar to the procedure outlined in the last section
for scalar metric fluctuations. For a mode with fixed comoving
wavenumber $k$, we compute the correlation function of the
off-diagonal spatial elements of the string gas energy-momentum
tensor at the time $t_i(k)$ when the mode exits the Hubble radius
and use (\ref{tensorexp}) to infer the amplitude of the power
spectrum of gravitational waves at that time. We then
follow the evolution of the gravitational wave power spectrum
on super-Hubble scales for $t > t_i(k)$ using the equations
of general relativistic perturbation theory.

The power spectrum of the tensor modes is given by (\ref{tensorexp}):
\be \label{tpower1}
P_h(k) \, = \, 16 \pi^2 G^2 k^{-4} k^3
\langle\delta T^i{}_j(k) \delta T^i{}_j(k)\rangle
\ee
for $i \neq j$. Note that the $k^3$ factor multiplying the momentum
space correlation function of $T^i{}_j$ gives the position space
correlation function, namely the root mean square fluctuation of
$T^i{}_j$ in a region of radius $R = k^{-1}$ (the reader who is
skeptical about this point is invited to check that the dimensions
work out properly). Thus,
\be \label{tpower2}
P_h(k) \, = \, 16 \pi^2 G^2 k^{-4} C^i{}_j{}^i{}_j(R) \, .
\ee
The correlation function $C^i{}_j{}^i{}_j$ on the right hand side
of the above equation was computed in Section 4. Inserting
the result (\ref{tensorresult}) into (\ref{tpower1}) we obtain
\be \label{tpower3}
P_h(k) \, \sim \, 16 \pi^2 G^2 {T \over {l_s^3}}
(1 - T/T_H) \ln^2{\left[\frac{1}{l_s^2 k^2}(1 -
T/T_H)\right]}\, ,
\ee
which, for temperatures close to the Hagedorn value reduces to
\be \label{tresult}
P_h(k) \, \sim \,
\left(\frac{l_{Pl}}{l_s}\right)^4 (1 -
T/T_H)\ln^2{\left[\frac{1}{l_s^2 k^2}(1 - T/T_H)\right]} \, .
\ee
This shows that the spectrum of tensor modes is - to a first
approximation, namely neglecting the logarithmic factor and
neglecting the $k$-dependence of $T(t_i(k))$ - scale-invariant. The
corrections to scale-invariance will be discussed at the end of this
section.

On super-Hubble scales, the amplitude $h$ of the gravitational waves
is constant. The wave oscillations freeze out when the wavelength
of the mode crosses the Hubble radius. As in the case of scalar metric
fluctuations, the waves are squeezed. Whereas the wave amplitude remains
constant, its time derivative decreases. Another way to see this
squeezing is to change variables to one
\be
\psi(\eta, {\bf x}) \, = \, a(\eta) h(\eta, {\bf x})
\ee
in terms of which the action has canonical kinetic term. The action
in terms of $\psi$ becomes
\be
S \, = \, {1 \over 2} \int d^4x \left( {\psi^{\prime}}^2 -
\psi_{,i}\psi_{,i} + {{a^{\prime \prime}} \over a} \psi^2 \right)
\ee
from which it immediately follows that on super-Hubble scales
$\psi \sim a$. This is the squeezing for gravitational waves.

Since there is no $k$-dependence in the squeezing factor, the
scale-invariance of the spectrum at the end of the Hagedorn phase
will lead to a scale-invariance of the spectrum at late times.

Note that in the case of string gas cosmology, the squeezing
factor $z(\eta)$ does not differ substantially from the
squeezing factor $a(\eta)$ for gravitational waves. In the
case of inflationary cosmology, $z(\eta)$ and $a(\eta)$
differ greatly during reheating, leading to a much larger
squeezing for scalar metric fluctuations, and hence to a
suppressed tensor to scalar ratio of fluctuations. In the
case of string gas cosmology, there is no difference in
squeezing between the scalar and the tensor modes. This supports
our expectation that the tensor to scalar ratio will be larger
in our scenario than in typical single field inflationary models.
We do find a relative suppression of tensor modes, but its origin
comes from string thermodynamics in the Hagedorn phase.

Let us return to the discussion of the spectrum of gravitational
waves. The result for the power spectrum is given in
(\ref{tresult}), and we mentioned that to a first approximation this
corresponds to a scale-invariant spectrum. As realized in
\cite{Ali2}, the logarithmic term and the $k$-dependence of
$T(t_i(k))$ both lead to a small blue-tilt of the spectrum. This
feature is characteristic of our scenario and cannot be reproduced
in slow-roll inflationary models.

To study the tilt of the tensor spectrum, we first have to keep in
mind that our calculations are only valid in the range (\ref{cond}),
i.e. to the large $k$ side of the zero of the logarithm. Thus, in
the range of validity of our analysis, the logarithmic factor
contributes an explicit blue tilt of the spectrum. The second source
of a blue tilt is the factor $1 - T(t_i(k)) / T_H$ multiplying the
logarithmic term in (\ref{tresult}). Since modes with larger values
of $k$ exit the Hubble radius at slightly later times $t_i(k)$, when
the temperature $T(t_i(k))$ is slightly lower, the factor will be
larger. Figure 3 shows the spectrum in the example of an
instantaneous transition between the Hagedorn phase and the
radiation-dominated period (i.e. no $k$-dependence of $T(t_i(k))$).

\begin{figure}
\includegraphics[height=5cm]{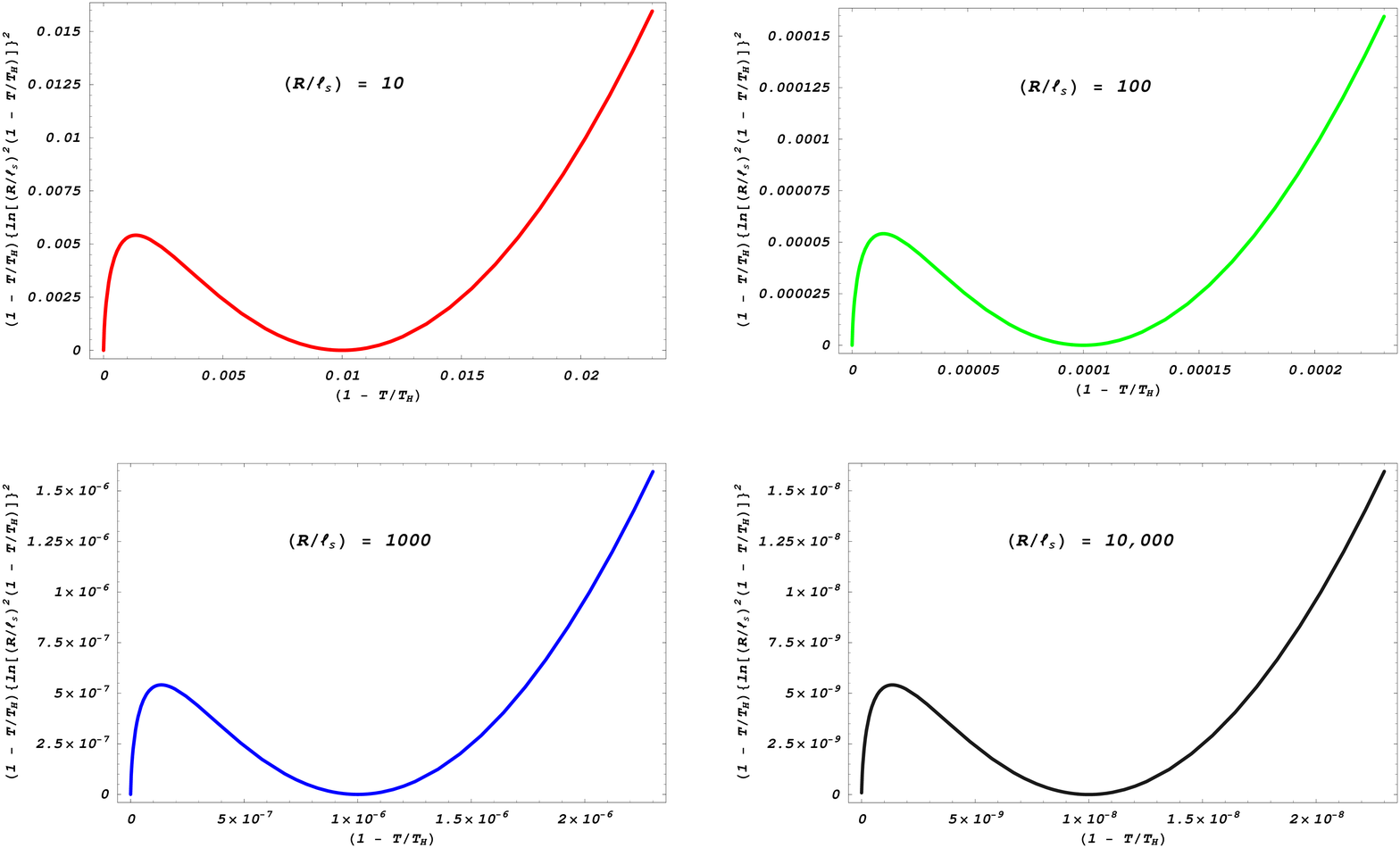}
\caption{Power spectrum of gravitational waves in string gas
cosmology in the case of an instantaneous transition between the
Hagedorn phase and the radiation-dominated period. In this
approximation, increasing the factor $1 - T/T_H$ is equivalent to
creasing $R$. The range of values of $R$ where our analysis applies
 to the left of the zero of the curve at the value $(l_s/R)^2$.
The increase of the power spectrum (the vertical axis) going towards
the left shows that the spectrum has a blue tilt.} \label{fig:3}
\end{figure}

Note that in the
result for the amplitude of the scalar metric fluctuations,
the prefactor $1 - T(t_i(k)) / T_H$ appears in the denominator
and hence leads to a slight red tilt of the spectrum. In addition,
the logarithmic factor is absent.

A heuristic way of understanding the origin of the slight blue tilt
in the spectrum of tensor modes
is as follows. The closer we get to the Hagedorn temperature, the
more the thermal bath is dominated by long string states, and thus
the smaller the pressure will be compared to the pressure of a pure
radiation bath. Since the pressure terms (strictly speaking the
anisotropic pressure terms) in the energy-momentum tensor are
responsible for the tensor modes, we conclude that the smaller the
value of the wavenumber $k$ (and thus the higher the temperature
$T(t_i(k))$ when the mode exits the Hubble radius, the lower the
amplitude of the tensor modes. In contrast, the scalar modes are
determined by the energy density, which increases at $T(t_i(k))$ as
$k$ decreases, leading to a slight red tilt.

\section{Discussion}

In this paper we have presented more details of the derivation
of the power spectra for scalar and tensor metric fluctuations.
Here we summarize the main results.

First of all, we have found that our string gas cosmology
background leads to a new way of obtaining a roughly scale-invariant
spectrum of cosmological perturbations. Our mechanism relies
on string theory in two crucial ways. Firstly, the cosmological
background is the result of a new symmetry of string theory which
is not present in ordinary quantum field theory, namely T-duality.
The background equations of motion which are consistent with the
T-duality symmetry, namely those of dilaton gravity, predict that
the radiation-dominated phase of standard cosmology was preceded
by a phase - the Hagedorn phase - during which the scale factor
in the string frame is static, the pressure vanishes, and the
temperature is close to the limiting temperature, the Hagedorn
temperature. As we have shown, string thermodynamical fluctuations
in the Hagedorn phase lead to an almost scale-invariant spectrum
of cosmological perturbations. This is the second point where
string theory plays a crucial role. Point particle fluctuations
would have given a completely different spectrum.

We have shown that the spectrum of the scalar metric fluctuations
has a slight red tilt, whose value depends on how fast the
transition between the Hagedorn phase and the radiation-dominated
period occurs.

Our model makes a unique prediction: the spectrum of gravitational
waves has a slight blue tilt, unlike the slight red tilt which is
predicted in scalar field-driven inflationary models. It would be
interesting to investigate how hard it would be to observationally
identify this signature. One factor which will help in this
regard is the fact that the tensor to scalar ratio $r$ in our
scenario is not suppressed by slow-roll parameters. As can be
seen immediately by comparing our results for the amplitude of
the tensor and scalar power spectra, respectively, the ratio $r$
at a scale $k$ is given by
\be r \, \sim \, (1 - T(t_i(k))/T_H)^2 \ln^2{\left[\frac{1}{l_s^2
k^2}(1 - T(t_i(k))/T_H)\right]}\, . \ee

In principle (if the dynamical evolution from the Hagedorn phase to
the radiation-dominated FRW phase were under complete analytical
control) this quantity would be calculable. If the string length were
known, the factor $(1 - T/T_H)$ could be determined from the
normalization of the power spectrum of scalar metric fluctuations.
Since the string length is expected to be a couple of orders larger
than the Planck length, the above factor does not need to be very
small. Thus, generically we seem to predict a ratio $r$ larger than
in simple slow-roll inflationary models.

We close by mentioning some limitations of our analysis. Since
the dilaton is evolving until the beginning of the radiation
phase, the gravitational perturbations should be determined and
evolved using the complete dilaton gravity background, not using
the equations of general relativity. The time dependence of the
dilaton will thus lead to correction terms. We are currently
in the process of calculating the magnitude of these effects.
This calculation will also be crucial in order to give a correct
estimate of the amplitude of the slight tilt in our spectra. These
calculations are also currently in progress.

We should also emphasize that, at least in the framework presented
in this paper, our cosmological scenario does not solve the
entropy and flatness problems of standard cosmology. In other words,
we need to start with three spatial dimensions which at the time $t_R$
are much larger than the string scale (the entropy problem), and
we can provide no explanation for why the current universe is so close
to being spatially flat. Inflationary cosmology, in addition to
providing a causal mechanism for structure formation, has the potential
of solving the entropy and flatness problems. On the other hand,
scalar field-driven inflationary cosmology is singular, whereas
string gas cosmology has the potential of providing a non-singular
cosmology.

\begin{acknowledgments}

The work of R.B. and S.P. is supported by funds from McGill University, by an
NSERC Discovery Grant and by the Canada Research Chairs program. The
work of A.N. and C.V. is supported in part by NSF grant PHY-0244821
and DMS-0244464.

\end{acknowledgments}

\end{document}